# Stakeholder Perspectives on Humanistic Implementation of Computer Perception in Healthcare: A Qualitative Study

Kristin M. Kostick-Quenet, PhD[1]; Meghan E. Hurley, MA[1]; Syed Ayaz, MA[1]; John Herrington, PhD[2]; Casey Zampella, PhD[2]; Julia Parish-Morris, PhD[2]; Birkan Tunç, PhD[2]; Gabriel Lázaro-Muñoz, JD, PhD[3]; J.S. Blumenthal-Barby, PhD[1]; and Eric A. Storch, PhD[1]

[1]Baylor College of Medicine, Houston, TX, 77030, USA

[2] Children's Hospital of Philadelphia, Philadelphia, PA 19104, USA

[3]Massachussetts General Hospital, Boston, MA 02114, USA

## Abstract

### Background
Computer perception (CP) technologies—including digital phenotyping, affective computing and related passivesensing- approaches—offer unprecedented opportunities to personalize healthcare, especially mental healthcare, yet they also provoke concerns about privacy, bias and the erosion of empathic, relationship-centered practice. A comprehensive understanding of perceived risks and benefits from those who design, deploy and experience these tools in real-world settings remains elusive.

### Objective

This study aimed to explore key stakeholder perspectives on potential benefits, risks and concerns around the integration of CP technologies into patient care. Better understanding of these concerns is crucial for responding to and mitigating such concerns via design implementation strategies that augment, rather than compromise, patient-centered and humanistic care and associated outcomes.

## Methods

We conducted in-depth, semi-structured interviews with 102 stakeholders spanning the CP life cycle: adolescent patients (n = 20) and their caregivers (n = 20), frontline clinicians (n = 20), technology developers (n = 21) and ethics, legal, policy or philosophy scholars (n = 21). Interviews (~ 45 min each) explored perceived benefits, risks and implementation challenges of CP in clinical care. Transcripts underwent thematic analysis by a multidisciplinary team; reliability was enhanced through double coding- and consensus adjudication.

## Results

Stakeholders articulated seven interlocking concern domains: (1) trustworthiness and data integrity; (2) patient-specific- relevance; (3) utility and workflow integration; (4) regulation and governance; (5) privacy and data protection; (6) direct and indirect patient harms; and (7) philosophical critiques of reductionism. A cross-cutting insight was the primacy of *context* and *subjective meaning* in determining whether CP outputs are clinically valid and actionable. Participants warned that without attention to these factors, algorithms risk mis-classification and relational harm. To operationalize humanistic safeguards, we propose "personalized roadmaps": codesigned plans that predetermine which metrics will be monitored-, how and when feedback is shared, thresholds for clinical action, and procedures for reconciling discrepancies between algorithmic inferences and lived experience. Roadmaps embed patient education, dynamic consent, and tailored feedback, thereby aligning CP deployment with patient autonomy, therapeutic alliance and ethical transparency.

## Conclusions

This multistakeholder study provides the first comprehensive, evidence-based account of relational, technical and governance challenges raised by CP tools in clinical care. By translating these insights into personalized roadmaps, we offer a practical framework for developers, clinicians and policymakers seeking to harness continuous behavioral data while preserving the humanistic core of care.

**Introduction**

Computer perception (CP) tools, including digital phenotyping, affective computing, computational behavioral analysis and other approaches that entail continuous and passive data collection using wearables and smartphone sensing, have been positioned as a remedy for longstanding diagnostic and therapeutic blind spots in mental healthcare. The term "computer perception" references the artificial intelligence (AI) subfield of computer "vision" but acknowledges a wider range of perceptive modalities beyond vision alone (e.g., "hearing" through microphones; motion detection through accelerometers, etc.). By leveraging sensors already embedded in everyday devices, these systems promise scalable, accessible surveillance of mood, cognition, and social behavior, potentially addressing the field's chronic reliance on infrequent patient self-reports and clinician observation [1,2]. CP tools also promise a personalized and patient-tailored diagnostic and therapeutic approach, in line with precision medicine goals [3, 4, 5]. Early studies suggest that CP-derived markers can forecast relapse in bipolar disorder, detect prodromal psychosis, tailor just-in-time behavioral prompts, and potentially widen access to mental health care. Yet the very features that make CP appealing also expose patients to unprecedented privacy risks, algorithmic bias, and the potential erosion of empathic, relationship-centered care [1, 6, 7, 8].

Ethicists, regulators, and frontline stakeholders caution that integrating such pervasive sensing into care can imperil core values of confidentiality, fairness, and relational trust [9, 10, 11, 12]. These impacts can be exacerbated by opaque algorithms, unclear pathways for secondary data reuse, and difficulties obtaining meaningful informed consent in continuous monitoring scenarios. A limited number of studies [13, 14, 15] provide a foundation for understanding some of these concerns; however, no empirical research to date offers a comprehensive view of the wide-ranging perspectives held by diverse stakeholder groups regarding the benefits and risks of integrating CP into care. The present study addresses this gap through an empirical exploration of diverse stakeholder perspectives, with special attention to impacts on humanistic, relationship-centered care.

The rationale for focusing on humanistic care is to balance enthusiasm for these tools with a recognition that healing is a human enterprise grounded in empathy, dignity, and shared decision-making. Humanistic and humanized care frameworks [16, 17, 18] remind us that respectful dialogue, cultural sensitivity, and patient partnership are interwoven into the moral fabric of good practice [19]. Whether CP ultimately augments or erodes that fabric depends on how well designers, clinicians, and regulators anticipate a spectrum of ethical concerns voiced by those who will build, deploy, or live with these systems. This study therefore turns to those diverse stakeholders – developers, clinicians, patients, caregivers, and ethics, legal and policy- scholars – to ask how their concerns can guide the integration of CP in ways that preserve, rather than diminish, the humanization of care.

## Background

What makes CP technologies unique is that they involve algorithmic *inferences* about a person's moment-to-moment mental, or socio-behavioral state or predicted outcomes such as mood relapse, suicidality, or treatment response [2, 12, 20, 21]. These inferences are enabled by the ingestion of vast amounts of behavioral, physiological, and environmental signals from (usually) ordinary connected devices such as smartphones and wearables. Less often, they may involve implantable systems to continuously monitor physiological [22, 23] or neural activity [24]. In psychiatric contexts, the approach is often called *digital phenotyping*, entailing the use of smartphones, wearables, and ambient sensors to stream accelerometry, GPS traces, keystroke dynamics, speech acoustics, heart rate variability, and other passively captured metadata. Those streams are preprocessed, feature-engineered [25, 26] and fed into statistical or deep learning models. Parallel work in *affective computing* [27] extends the approach to facial micro-expressions, vocal prosody, or text sentiment to classify discrete emotions or arousal levels in real time. [59] extends the approach to facial micro-expressions, vocal prosody, or text sentiment to classify discrete emotions or arousal levels in real time.

Because CP systems sit at the intersection of pervasive sensing and advancements in AI, they raise many of the same ethical issues highlight by AI systems. Concerns about algorithmic bias, transparency, explainability, interpretability, fairness and other aspects of "trustworthy" AI [28, 29] are relevant. The rarity by which CP tools are validated on large, diverse validation cohorts means that algorithmic performance is likely to vary dramatically across demographic and clinical groups, raising reliability concerns and potentially amplifying health disparities [30]. Critics have also warned against over-reliance on algorithmic inferences about patients' health status [31, 32], especially "black box" systems that resist clinical scrutiny and accountability and compromise informed clinical decision-making [33]. Others [15,34] underscore legal uncertainties surrounding liability in cases of error, patient harm, or mismanagement of outputs or other feedback. The U.S. National Institute of Standards and Technology's AI Risk Management Framework and the EU AI Act both categorize health-related CP tools as "high risk," demanding rigorous safety, fairness, and oversight provisions [35, 36].

Similar to other AI systems, CP tools thrive on voluminous data sets, not only across individuals but for each individual, often called individual "big data" or "deep data" [37]. Ethical critiques thus consistently foreground privacy vulnerabilities associated with sensitive behavioral data [13, 14, 38, 39], and there is expert consensus [13] around the need for privacy and innovative consent approaches. Scholars [34, 40, 41] caution against unwanted or involuntary disclosure to third parties, such as insurers or employers, in scenarios where data are controlled by consumer-grade device companies. Dynamic consent models have also been proposed [39] to replace onetime or broad consent approaches with ongoing, granular permissions; however, feasibility remains challenging [42].

### Challenges for Humanistic Care

Critics [9, 43, 44] have also converged on a deeper worry: as algorithms assume a larger share of the responsibility to observe and listen, the relational core of care risks being reduced to a "metrics management" exercise, whereby clinicians and patients spend their limited time consulting data trends rather than discussing the patient's lived experience and therapeutic goals. Clinicians fear that multimodal dashboards could displace narrative dialogue, shifting the burden of self-monitoring and, by extension, responsibility for changes in functioning, onto patients in ways that compromise dignity and mutual trust [9, 45-47] and overprioritize technological over humanistic solutions [48]. Some warn that automated detection and treatment of illness may weaken the rapport and goal alignment that bolsters the therapeutic alliance, unless paired with explicit, empathic communication strategies [49].

A limited set of empirical work reinforces these cautions. One study [47] documents mental health clinician enthusiasm toward gaining rich, real-time insights but also highlights concerns about workflow overload and potential for automation bias, i.e., deferring to algorithmic judgments even when they conflict with a clinician's intuitions or a patient's lived story. Another study [46] highlighted clinicians' concerns that prioritizing passive data trends over self-reported narratives or active responses to clinical assessments could reduce opportunities for patients to reflect upon their mental health, leading to reduced patient engagement. Experts [9, 21] have raised flags that such asymmetries can tilt encounters toward dehumanization and require careful planning and implementation to achieve the goal of making otherwise invisible patterns visible and clinically useful.

These relational stakes bring long-standing ethical principles into focus and urge clinicians and researchers to keep dignity, empathy, patient empowerment, and shared decision-making at the forefront of clinical care. However, it remains- unclear how best to do this in ways that engage multiple and often competing perspectives. Our study addresses this gap by exploring the range of concerns through interviews with over 100 stakeholders across the CP lifecycle. We catalogue considerations that extend beyond well-elaborated privacy and bias debates to the less-operationalized relational harms that data-centric care may impose. By situating these concerns within established humanistic frameworks of dignity, empathy, and shared decision-making [17, 18], we offer an anticipatory roadmap for researchers, developers, clinicians and patients. The goal is not merely to identify technical fixes, but to ensure that as CP systems mature, they deepen rather than diminish the person-centered relationships that remain the centerpiece of care.

**Methods**

As part of 4-year study funded by the National Center for Advancing Translational Sciences (R01TR004243), we conducted in-depth, semi-structured interviews (total n=102), including with adolescent patients (n=20) and caregiver (n=20) dyads, clinicians (n=20), and developers (n=21), and ethics, legal, policy and philosophy scholars (n=21) to explore their perspectives on potential benefits, risks and concerns around the integration of CP technologies into their care. Respondents were recruited from a "sister" study (5R01MH125958) aiming to validate CP tools designed to quantify objective digital biobehavioral markers of socio-emotional functioning.

**Participants**. Participants included a clinical sample of adolescents (aged 12-17 years) with varied diagnoses, including Autism, Tourette's, Anxiety, Obsessive-Compulsive Disorder (OCD), and Attention Deficit/Hyperactivity Disorder (ADHD), as well as their caregivers (typically biological parents, Table 1). Diagnostic presentations for all adolescents were confirmed by expert providers using established clinical measures. Adolescent-caregiver dyads were referred to the current study by the sister study's coordinator and then contacted by a research assistant via phone or email to schedule an interview. Clinicians and developers (Table 2) were identified through online literature search and through existing professional networks. Participants were interviewed between January 2023 and August 2023.

| Table 1. Demographics for Interviewed Adolescents and Caregivers | | | Adolescents | | Caregivers | | TOTAL |
|---|---|---|---|---|---|---|---|
| | | | n = 20 | % Total 50% | n = 20 | % Total 50% | n=40 |
| Gender | | Male | 12 | 30% | 2 | 5% | **35%** |
| | | Female | 8 | 20% | 18 | 45% | **65%** |
| Ethnicity | | American Indian or Alaska Native | 0 | 0% | 1 | 3% | **3%** |
| | | Asian | 1 | 3% | 1 | 3% | **5%** |
| | | AA/Black | 5 | 13% | 4 | 10% | **23%** |
| | | Native Hawaiian or Other Pacific Islander | 0 | 0% | 0 | 0% | **0%** |
| | | White | 17 | 43% | 15 | 37% | **80%** |
| | | Hispanic or Latino | 4 | 10% | 2 | 5% | **15%** |
| | | Not Hispanic or Latino | 16 | 40% | 18 | 45% | **85%** |
| Marital Status | | Married and living with spouse | - | - | 13 | 65% | **65%** |
| | | Widowed | - | - | 1 | 5% | **5%** |
| | | Divorced | - | - | 4 | 20% | **20%** |
| | | Separated | - | - | 1 | 5% | **5%** |
| | | Never Married | - | - | 1 | 5% | **5%** |
| | | High school only or less | - | - | 0 | 0% | **0%** |
| | | Trade school/Associate's degree | - | - | 2 | 10% | **10%** |
| | | Bachelor's degree | - | - | 10 | 50% | **50%** |
| | | Master's degree | - | - | 4 | 20% | **20%** |
| | | Doctoral degree | - | - | | | |

| | | Clinicians n | Clinicians % | Developers n | Developers % | TOTAL |
|---|---|---|---|---|---|---|
| | | - | - | 4 | 20% | **20%** |
| Parental Status | Biological Parent | - | - | 18 | 90% | **90%** |
| | Step Parent | - | - | 0 | 0% | **0%** |
| | Adoptive Parent | - | - | 2 | 10% | **10%** |
| Diagnosed Condition | OCD | 4 | 20% | - | - | **20%** |
| | Autism | 5 | 25% | - | - | **25%** |
| | ADHD | 3 | 15% | - | - | **15%** |
| | Anxiety | 4 (1 self- reported) | 20% | - | - | **20%** |
| | Tourette's | | | | | |
| | No clinical diagnosis or symptoms | 1 | 5% | - | - | **5%** |
| | | 9 | 45% | - | - | **40%** |
| | Average age | 14.9 | (s.d. 2.2) | 48.3 | (s.d. 6.4) | |
| *\* Values may not total to 100% due to categorical overlap (e.g., comorbidities).* | | | | | | |

**Table 2. Demographics for Interviewed Clinicians, Developers, and ELPP**

| | | Clinicians | | Developers | | Scholars | | TOTAL |
|---|---|---|---|---|---|---|---|---|
| | | n = | % Total | n = | % Total | n = | % Total | |
| | | 20 | 32% | 21 | 34% | 21 | 34% | **n=62** |
| Gender | Male | 10 | 50% | 18 | | 16 | | **55%** |
| | Female | 10 | 50% | 3 | | 5 | | **29%** |
| Profession | Clinician | 3 | 15% | | | | | **5%** |
| | Clinician-Researcher | 14 | 70% | | | | | **23%** |
| | Clinician-Developer | 3 | 15% | 4 | 19% | | | **11%** |
| | Developer | | | 17 | 81% | | | **34%** |
| | Ethicist | | | | | 6 | 29% | **10%** |
| | Lawyer | | | | | 4 | 19% | **6%** |
| | Philosopher | | | | | 1 | 5% | **2%** |
| | Other | | | | | 10 | 48% | **16%** |
| Specialty | Psychiatry | 7 | 35% | | | | | **11%** |
| | Psychology | 7 | 35% | | | | | **11%** |
| | Neuroscience | 4 | 20% | | | | | **6%** |
| | Industry | | | 15 | 71% | | | **24%** |
| | Academic | | | 3 | 14% | | | **5%** |
| | Cross-Sector | | | 3 | 14% | | | **5%** |
| | Ethics | | | | | 6 | 29% | **10%** |
| | Law | | | | | 4 | 19% | **6%** |
| | Philosophy | | | | | 1 | 5% | **2%** |
| | Other | 2 | 10% | | | 10 | 48% | **16%** |

**Data Collection**. Separate but parallel interview guides were developed for all stakeholders, with the same constructs explored across groups, including: perceived benefits and concerns regarding integrating CP tools into clinical care, impacts on care, attitudes towards automatic and passive detection of emotional and behavioral states, perceived accuracy and potential for misinterpretation/-attribution/-classification of symptoms or conditions, clinical utility and actionability, data security and privacy concerns, potential for unintended uses, perceived generalizability and potential for bias, and other emergent concerns. These domains were chosen based on issues raised in the clinical and ethics literatures and with the guidance of experienced bioethicists and mental health experts. Initial drafts of the interview guides were piloted with two psychologists (ES, CJZ) specializing in adolescent mental health, resulting in minor clarifications in wording. Interviews were conducted via a secure video conferencing platform (Zoom for Healthcare) and lasted an average of ~45 minutes. This study was reviewed and approved by the [redacted] Institutional Review Board (H-52227), which waived a requirement for written consent; thus, participants provided verbal consent.

**Data Analysis**. Interviews were audio-recorded, transcribed verbatim, and analyzed using MAXQDA software. Led by a qualitative methods expert (KK-Q), team members developed a codebook to identify thematic patterns in adolescent and caregiver responses to questions addressing the topics above. Each interview was coded by merging work from two separate coders to reduce interpretability bias and enhance reliability. We used Thematic Content Analysis [50, 51] to inductively identify themes by progressively abstracting relevant quotes, a process that entails reading every quotation to which a given code was attributed, paraphrasing each quotation (primary abstraction) and further identifying which constructs were addressed by each quotation (secondary abstraction), and organizing constructs into themes. To enhance the validity of our findings, all abstractions were validated by at least one other member of the research team. In rare cases where abstractions reflected different interpretations, members of our research team met to reach consensus.

## Results

Stakeholders raised a wide range of concerns around the following themes: 1) Trustworthiness and Integrity of CP Technologies (Table 3); 2) Patient-Specific Relevance (Table 4); 3) Utility and Implementation Challenges (Table 5); 4) Regulation and Governance (Table 6); 5) Data Privacy and Protection (Table 7); 6) Patient Harms (Table 8); and 7) Philosophical Critiques (Table 9). All themes and sub-themes are elaborated below, with illustrative quotations in the associated Tables 3-9.

### *Trustworthiness and Integrity of CP Technologies*

**Data Quality Constraints and Confounds.** Developers, more than other groups, raised concerns about the reliability of data streams from consumer-grade devices, noting that variations in user behavior and differences in hardware performance can make it difficult to distinguish true physiological changes from sensor errors. They cautioned that without standardized protocols for device calibration and data collection, models built on these inputs risk failing when deployed across different environments or patient populations.

**Algorithmic Bias and Generalizability.** Participants from all groups also expressed concern over other types of algorithmic bias. Some scholars emphasized that many AI models are trained on homogenous datasets, limiting their applicability to broader, more diverse groups. They explained that because these algorithms often derive from data collected predominantly in privileged populations (e.g., younger, healthier, or majority-ethnic cohorts), they may underperform or misclassify signals in marginalized communities. Furthermore, they raised concerns that unequal access to digital health tools skews training datasets even further, reinforcing systemic biases and potentially excluding those most in need.

**Construct Validity.** Clinicians, developers and scholars alike warned that the underlying diagnostic constructs and clinical assessment tools used to validate most CP tools lack strong links to clinically meaningful constructs, and cannot accommodate transdiagnostic symptom manifestations, cultural and contextual variation, and temporal fluctuations in mental health; thus, the resulting digital markers of disease remain ungrounded. They emphasized the necessity of rigorous validation studies to ensure that digital biomarkers accurately reflect patient states and that any interventions based on these measures are grounded in well-established clinical evidence.

| Table 3. Accuracy, Validity and Trustworthiness of CP Tools | | |
|---|---|---|
| Themes | | Illustrative Quotes |
| **Data Quality Constraints & Confounds** | Behavioral Variability & Consistency | **You have to control for things that are easily overlooked,** like the fact that **people don't wear a wearable all the time**. In fact, we know at the population level that about **half of the population stops wearing a wearable after four to five months. And, 90% stop wearing wearable after one year.** People who are **typically most compliant are typically young, healthy individuals,** especially people who are really into sports and exercise and stuff like that." (D_06) |
| | | "Because if everything is based on that, the model accuracy, then decision-making and everything. **If you wear your watch loosely, you don't get good measurements.** And then the other parts are if you have a software update you miss. Software updates all the time, like **the robustness to all those external factors, which you're not controlling. I feel like it is the biggest set of problems in my opinion, operational problems.** So in research protocols you can manage to reduce these compounding variables, but once you deploy a system, this will become universal issues. And I feel like that will be a major bottleneck for |
| | | widespread use, which is people can't get reliable data first." (D_18) |
| | Device Variability and Performance | "Of course there's the **hardware element of this as well in terms of light sensitivity to darker skin**, all that kind of thing" (ELPP_07) |
| | | One challenge here is that these **commercial wearables don't tell you when the device is worn by the person. But, research grade devices do...** But when [validation] is done outside of a lab... in free living settings, that's actually when you have some of these big challenges. (D_06) |
| | Generalizability | I remember reading that s**ome groups of people, when they speak, they don't move their body a lot,** so they don't have a lot of body language. **So for the type of machine that reads body language, I wouldn't really be effective with them**. (P_11) |
| | | Only 5% of the models in individualized clinical prediction models in psychiatry … get externally validated, which means that 95% are not generalizable." (C_09) |
| | | "If you use them **in a context where you have very little data, they will overfit. A**nd, now you have a true **problem with generalizability.**" (D_06) |
| **Algorithmic Bias** | Non-diversity of training data sets & developers | But I feel like it **could easily become something where it can marginalize a group and not give a certain group of people the right care because it misunderstands something or it's created by a certain race of people.** And then, it only applies to that certain race of people... Or gender might play a role in it. **If it's created by men, then women may not be able to use it as efficiently.** (P_07) |
| | | "I think there's **not very good data that we have in general on various ethnic populations**. So, I worry about that. I think we're making a lot of generalizations, and I know that the likelihood of some populations to be able to give data, just access to the population is limited. And so, **I worry about making leaps for whether it's a minority person or an age group that's hard to reach, older, younger, and anyone whose data is not there on...**" (C_08) |
| | | "**Who has access to this tool?**… That's **going to create a bias in the data sets that we get** and the people that have access to this type of care, those are our potential concerns as well." (ELPP_09) |
| | | "It's really **easy to get convenience samples that are primarily white and primarily high SES families**, the kinds of families who can take a day off work to come in for a research study. And **that is not the average child on the spectrum. And so I do have big concerns that computer vision folks will not sufficiently attend to the kinds of** |

| | | | |
|---|---|---|---|
| | | | variables that could really impact things." (C_19) |
| | | | "I see a lot of stuff about **how poorly trained in terms of demographic groups, the facial affect technology is.** So maybe it works great on me because I'm a middle-aged white guy and there's lots of pictures of me. But if it thinks that Black faces are angry because it was trained on mug shots, and there's more people arrested who are Black, not because they commit more crimes, but because that's where the police are, then **the system is just eating its tail, creating all kinds of perpetuating of injustice... The data on which things like the facial computing ethics stuff is trained [is] going to have problems like tha**t." (D_08) |
| | | | "**The issue was that was the training stage.** That there was hand labeling done beforehand and then the AI system will only do what it was trained to do in the first place. So **that bias issue is very much present in emotion based questions as well."** (ELPP_07) |
| | Off-Label Use | | "…you c**an think of many different use cases… [for]** algorithms **for particular populations** or **for purposes for which they were not initially trained or intended.**" (ELPP_17) |
| | Variability in Symptom Expression | | "There are these **intangible aspects…** cultural, historical aspects of how we think about emotions that **don't necessarily get reflected in these model building…**" (ELPP_17) |
| | | | "My brain immediately goes to facial recognition technology that's used in criminal legal systems and how bias is so deeply baked into that. And I'm thinking about **the cultural constraints of affective expression and gesture...** I see a lot of first generation kids whose parents are refugees, or immigrants in their adulthood, and **I don't even know enough about how acculturation impacts affective expression. I think that's a concern for me..."** (C_13) |
| | | | "How can you really control for the fact that **a smile might mean something in different contexts?** So not only within cultures, but across cultures…" (D_03) |
| **Construct Validity** | Validity of Existing Diagnostic Categories and Assessments | Uncertainty around Training Targets and Ground Truths | "A lot of **this tech kind of assumes that the diagnostic tools we have are cross-cultural...** But... Those categories **might not be completely true."** (ELPP_01) |
| | | | "You **might inadvertently...create...novel kinds of clusters** that **do not necessarily map onto our preexisting conceptual understandings** of categories" (ELPP_18) |
| | | | "**What are we training to?** Are we training to a PHQ-9? Are we training to a Hamilton? Are we training to a clinical diagnosis or training the DSM? **It is not really clear to me. I think that's still been the biggest handicap for this field...**" (ELPP_20) |
| | | | "How do we account for even the fact that disorders that we are trying to detect discreetly and separately from each other, might actually be... Like problems of living, they might be **network problems rather than sort of distinct entities that can be detected and discerned.**" (ELPP_01) |
| | | Uncertain Illness Ontologies | "…I think the biggest thing that scares me… we **don't really have any objective markers**… we're **kind of assuming that there's an objective entity that we can find**, and that data's going to be the answer…" (ELPP_16) |
| | | | ; "...part of what I think **has been so problematic for mental health is the diagnostic scheme is A, doesn't really have a strong scientific basis using any kind of objective measures** and B, link in any way to treatment response or etiology, either one of those. So in some ways, it's a harder task than it is when you're looking for digital phenotyping in some other areas of medicine. And I guess the best example is using AI to read mammograms where you have a ground truth from having done a biopsy and which tumors are malignant an**d which are not. We just don't have that here. We don't have a biopsy, we don't have any solid footing that we're going to go up against, and that to me is the biggest challenge for the field.**" (ELPP_20) |
| | External Cues May Not Reflect Internal Experience | Behavioral Expressions | "I think in terms of the idea of reverse inference, so **you have an outward signal to detect an interior state...** So from a clinical side of things, you have problems of diagnosis, is **what you're sensing from the outward. Does that have anything to do with what's going on inside a person?**... I think [**the assumption**] that [**there's**] a **connection between internal and external expression that's forever lasting and reliable is problematic,** I think." (ELPP_07) |
| | | Physiological (Bio-) Markers | "I think [physiological changes] can tell us something about the internal state of a patient... Is the heart rate spiking? What is heart rate variability? What is the skin conductance? How much are you sweating? You can tell all these things. [But] **I don't think you can use that data to then tell us something about what a person is feeling.**" (ELPP_01) |
| **Explainability** | | | **If the model doesn't do what it's supposed to do, you can't open the box and say, "Ah, this is why it doesn't work."** (D_06) |

*Patient-Specific Relevance*

**Accounting for Heterogeneity in Symptom Expression & Subjectivity.** Stakeholders consistently emphasized that any use of digital health tools must first reckon with the immense diversity in how individuals experience and express their health and then situate those signals within each person's unique context. Respondents from all groups warned that a one-size-fits-all algorithm may miss or misinterpret patients who manifest emotional or behavioral states differently than others; for example, some pointed out that certain individuals express distress outwardly while others internalize such feelings, rendering them "invisible" to CP tools searching for external markers. Others added that accurate interpretation often depends on integrating multiple data streams; heart rate alone, for example, may not distinguish stress from exercise without knowing the broader context or behavioral pattern.

**Accounting for Context & Meaning**. Patient and caregivers, more than other groups, raised concerns that algorithms cannot effectively account for the rich social and cultural factors that shape patient's experiences and behaviors, or how patients assign meaning to their symptoms and events. Some also highlighted the importance of proximate contextual features, such as fluctuations tied to work demands, family stressors, or lifestyle changes. Patients in particular cited concerns that algorithms might come to conclusions based on fleeting or temporary signals rather than longer-term trends. Respondents across groups warned that such "decontextualized" metrics lack the construct validity required for clinical actionability, as they are likely to represent inferences stripped of subjective meaning and, therefore, clinical significance.

**Table 4. Patient-Specific Relevance**

| Themes | | Illustrative Quotes |
|---|---|---|
| Accounting for Heterogeneity in Symptom Expression & Subjectivity | Capturing Subjective Meaning and Significance | "I'm a very type A person. Go, go, go. To stay home for a while or don't get out of bed... even if I'm really stressed or having a bad day, I don't do that, because I'll always have a to-do list. Even if I'm having a bad day, I have to do these things. **[The system] might not realize sometimes that I'm overly stressed or upset, just because I'm still going on with my normal day."** (P_17) |
| | | "Even objective data needs confirmation from subjective reports, need to understand what patients were going through internally when data was collected - have good quotes on this; "I think [these technologies would be useless without subjective input] in the sense that there's no good healthcare without patient consent and input because it's about their health. **It's not about what we think is going on with their health, it's about their experience of it.** We should not use it if we are not going to take what the patient says into consideration." (C_12) |
| | Capturing Individual heterogeneity | "There's a **huge amount of individual variability**, and I think **one of the things we learned is this is harder than it looks."** (ELPP_20) |
| | | S"o, I think some people are very emotive and other people ... For example, for some people when they become anxious, they're nervous, they're fidgety, they're sweating, they're tremulous... But there are some people when they become anxious, they look disinterested... not tuned in... not paying attention. So **people will express those emotions differently. And would a computer be able to pick that up?"** (CG_01) |
| | | "If I'm sad, it's pretty obvious, if I'm pissed off, it's pretty obvious, but I know a lot of other people it would probably be harder. I**s [the algorithm] going to be able to pick up on the same exact signs it would for someone else?"** (P_19) |
| Accounting for Context & Meaning | Sociocultural and Environmental Context | **"That behavior has to be understood in context**… It **doesn't necessarily mean that the algorithm's going to work well in a very different context**. And that's an empirical question." (D_03) |
| | | "[With wearable technologies,] **all that context gets lost.** And that **context is almost always informed by what is your life constellation?** What's your **cultural background?** Your **socioeconomic situation? Are you rural or urban?** And, and, and…" (D_20) |
| | | "Yeah, well, it's not [personalized medicine]. It **doesn't account for all really important things that happen in my day-to-day life. They don't appreciate my culture,** necessarily. They don't understand the sort of social issues I'm having, or they don't understand just these day-to-day things are going on in life." (ELPP_14) |
| | Personal Attributions of Meaning | "It's **the context piece**, that you can see changes relative to context. And **context is often a very personal and private thing.**" (C_04) |

| | | "Maybe **if something in particular happened that made you sad** for a period of time but **it's not permanent,** then I know that **I wouldn't want that to be taken out of proportions and maybe it would become a 'thing', even if I didn't want it to, it was just something temporary...**stuff like actual emotions, I don't know... **sometimes those change really fast and I feel different... I don't know if it would be able to pick up on that as well as it thinks it can."** (P_14) |
| Temporal Contexts | | "I think for the most part I'd be fine with [passive data collection], but **if it's constantly picking up on the five minutes that I just didn't get exactly what I wanted** and it **would just be like, 'Oh my gosh, you're not feeling okay, you are possibly depressed,'** and it's like, **'No, it was just a five minute thing.'** I feel like it **could eventually get to the point where it's annoying**..." (P_19) |

*Utility and Implementation Challenges*

**Role of CP in Clinical Care**. Stakeholders from all groups voiced a set of interrelated concerns about how CP tools are integrated into clinical workflows. Scholars and clinicians cautioned that clinicians may lean too heavily on algorithmic outputs, risking a form of "deskilling" in which they stop rigorously scrutinizing the data for quality or epistemic inconsistencies. They warned that clinicians may come to accept CP suggestions uncritically (automation bias), thereby sidelining the human, relational interpretations arrived at through patient–provider dialogue.

**Managing Risk and Liability.** Clinicians, more than other groups, highlighted the dual dangers of missed events and over-alerting. They noted that false negatives – instances where the system fails to detect deterioration – could leave patients unprotected, while excessive false positives could overwhelm clinicians and erode confidence in the tool, ultimately undermining patient safety rather than enhancing it. Clinicians also raised concerns about whether they may eventually be expected to use CP tools as they continue to evolve, or held liable if they choose not to, compromising their autonomy in clinical decision making.

**Barriers to Utility.** All stakeholder groups stressed that CP outputs must be interpretable and meaningful in real-world contexts to be actionable. Clinicians stressed that data trends and inferences must be delivered through intuitive summaries and visualizations, accompanied by concise, actionable recommendations. They described how this is complicated by observations that the clinical significance of data trends is likely to vary from one situation to another (see Patient Specific Relevance above), complicating reliable interpretation. Developers and clinicians also expressed concerns about the potential for confirmation bias, where users may cherry-pick data that confirms their expectations, undermining the aim of these technologies to provide novel informational value to clinical assessments.

| Table 5. Utility & Implementation Challenges | | | |
|---|---|---|---|
| Themes | | | Illustrative Quotes |
| **Role of CP in Clinical Care** | Tech Overreliance | Clinician Deskilling | "And there is research about **clinicians relying too much on the decision support and not questioning it.** A bit like somebody on a conveyor looking at a conveyor belt in a manufacturing factory where there was AI sensing for defects. **In the end they're not doing anything. They're like someone sitting in a driverless car. They're not really a human in the loop...** Who's making life changing decisions or life affecting decisions?" (ELPP_04) |
| | | | "And so I think what you're seeing is that there may be a distributional gulf that increases in the quality of hyper users with maintaining the domain expertise and low intensity users who become in fact fully reliant on the systems." (D_19) |
| | | | "In the worst case scenario, I was thinking about how would I feel about this if I was a first year therapist in grad school doing training, and I wonder... It could be that it's useful because it's like, 'Oh, here's some things to consider.' But it also **could be that it makes someone doubt their own burgeoning clinical judgments or, 'Well, the computer said this, so I guess I must have missed that and let's go with that.'"** (C_15) |
| | | Automation Bias | "If an AI just **presents one recommendation,** then the **user could be prone to automation bias.** Automation bias being the over-reliance or the **complacency of just accepting whatever the AI says**, without... And are we as rigorous in looking at the data when there's AI assisting?" (D_15) |
| | | | "If there's some sort of problem with [the inferences] and you're just sort of taking it as like, "Oh, okay, this is what it says", **it could make you make the wrong sorts of decisions. Or again, maybe make you think... 'Maybe they** *do* **have a personality thing, I wasn't detecting that, but maybe they do.' And maybe you just sort of muddle the picture up a little bit in the worst case scenario."** (C_15) |
| | | Loss of Human/Relational Interpretation | "AI is good at picking out patterns… but **quite bad at understanding what that pattern means in the context of the individual**… it's still really about the clinician and their patient." (C_04) |
| | | Managing Risk and Liability | "Then, obviously, the false negative too, right? I mean, as a clinician, **I worry about, say, there were an adverse outcome,** following what looked like normal readings from the technology. **Does that put me in the hook for litigations and so on?"** (C_16) |
| | | | "And I think **other risks are also the liability**. There's a reason why, when we automate depression screeners, we exclude the question that asked about suicidality, because **there's an obligation to then take that seriously... what would happen if the system isn't able to respond in the moment to that?"** (C_13) |
| **Barriers to Utility** | Interpretability | Consistency of Meaning | "If you're a company and your whole business model is based on sending out alerts when somebody's having a psychiatric episode, **it has to be based on a real finding, on a real phenomenon...** If you roll it out at scale, **it has to be something that's repeatable. [But] making sense of the data is a lot more difficult than people realize... It's harder to extract the gold than what people really think."** (D06) |
| | | | "People don't understand. **Let's say I make a decision for a patient on software version 1.0.3.1, and then somewhere in the background it got updated. Now, the algorithm says I should have made a different decision, or it shifts the probabilities in my differential diagnosis differently.... Now what do I do in this case? There's a legal issue, there is a care issue.** It's not like a thermometer, which is |
| | | | going to be thermometer yesterday and today the same way. **An algorithm** doesn't behave that way; it **has a fluidity... which we do not know... how to really deal with it.** (D_18) |
| | | | "Even if you trust the heart rate variability numbers that come out of this actigraph that you're using, then what do they correlate to? **Do they always mean this? Do they always mean that? How do we know that it means that in this population or that population?** What is the practical effect really?" (C_20) |
| | | Confirmation Bias | "In radiology specifically and diagnostic imaging fields, **the clinicians' ability is they're experts in interpreting images and being able to communicate that knowledge...** We know that clinicians like radiologists have training in being able to identify whether there's a pulmonary nodule or not, **but I think we need to recognize that the introduction of these (CP) models, if not designed carefully, can lead to confirmation bias, which is looking at things in the image to confirm what you want to see in the image.** [Conversely], **if you're someone, you do not consider yourself at an expert level, let's say in training, you will be in a cognitive dissonance stage where you are questioning yourself and believing that the model is somehow truthful,** and that can lead to more inefficiencies than efficiency if this happens a lot." (ELPP_10) |
| | | Data Overload | "If I was recording all my weight and GSR and if I was recording EEG during sleep, pulse oximetry at home and blood pressure, and... uploading that into some electronic care record, will my GP feel responsible to read that and interpret that? **That kind of burden of lots and lots of information and data.. The more signals that we're providing, then they need to be interpreted. But is it by a computer, by a human or both?**" (D_15) |
| | | | "I think that what will probably happen on the clinical side is that people will get into wearables and these things, patients themselves. And, it **may be difficult for clinicians to make sense of the data or these metrics** because it might be a different device, it could be something very complicated. So, I could see on the clinician side that **clinicians might feel overwhelmed**." (D_06) |

| | | |
|---|---|---|
| | Risk of Non-Actionable Metrics | "A six-minute walk test **might seem like a meaningful measure to a patient** going into a clinical evaluation for how well they ambulate, but we've seen over the course of some decades that **it's not actually meaningful.**" (ELPP_19) |
| | | "...even if you trust the heart rate variability numbers that come out of this actigraph that you're using, then what do they correlate to? Do they always mean this? Do they always mean that? How do we know that it means that in this population or that population? **What is the practical effect really?**" (C_20) |
| | False Alarms / Positives | "Having the measure is never a problem. It's **how it gets used and who it's provided to, that can create the problem.** And so, people can be falsely alarmed by these kind of things." (D_13) |

## Regulation and Governance

**Unclear or Insufficient Regulatory Frameworks.** Clinicians and scholars, more than other groups, described two distinct but related regulatory challenges. First, many CP applications can (and in their view, should) fall under existing clinical use regulations, such as those governing medical devices; yet few concrete guidelines exist on how to implement these requirements in practice. Ethics and policy experts noted that when CP tools nominally qualify as regulated devices, organizations may feel more comfortable adopting them; however, the absence of clear, step-by-step governance pathways often leaves developers and clinicians wondering how to operationalize data privacy, security, and ethical review processes. Second, participants emphasized that a large swath of CP technologies occupies a "regulatory grey zone" due to their overlap with many devices employed in the consumer "wellness" sector, particularly those that collect passive or contextual data outside traditional care encounters. Scholars worried that by failing to specify oversight requirements for continuous, ambient monitoring, regulators risk leaving patients exposed to unvetted algorithms and unclear lines of accountability.

**Responsibility for Ethical Tech Development and Compliance**. Developers, scholars, and clinicians, primarily, expressed concerns about how innovation pressures interact with ethical safeguards. On the one hand, experts described the burden of balancing innovation against regulatory demands, reporting that small teams sometimes struggle to absorb the time and cost required for formal ethics and security reviews. They also raised concerns about deployment of closed-source, proprietary algorithms, which are often faster to market but opaque. They contrasted these with opensource alternatives, which permit external audit but come with greater technical support obligations. Across both choices, questions about liability were raised, with respondents arguing that without explicit legal clarity, neither developers nor healthcare providers know with certainty who would be held accountable if CP assessments lead to harm.

**Need for Stakeholder Involvement.** Respondents from all groups expressed strong consensus that regulation and governance structures must be codesigned with the people intended to benefit from these technologies. Ethics scholars argued that embedding patients' and caregivers' lived experiences into standards-setting is vital to ensure tools address real-world needs. Clinicians highlighted the importance of rigorously interrogating when and under what circumstances CP outputs truly matter to patient care, rather than assuming technological assessments will always be relevant. Moreover, participants from all groups called for interdisciplinary collaboration among technologists, clinicians, ethicists, and end-users to bridge gaps in expertise, surface hidden risks, and develop governance models that are both practical and ethically robust.

**Table 6. Regulation & Governance of CP Technologies**

| Themes | | Illustrative Quotes |
|---|---|---|
| **Acceptability & Uptake Requires Regulation** | Subject to Existing Clinical Use Regulations | "One thing that shifts the calculus a lot is that the **technologies that are going to get used in a clinical setting... will be subject to more regulation than those that are [direct-to-consumer]...** I think the regulation, as much as people hate regulation, I think is a ...key factor here because... whether you call it consumer use or patient use, [what] **you're engaged in essentially medical interactions or interventions.** So I think that's really different, the regulation part." (ELPP_12) |
| | | "Many of the companies that are creating these technologies are **not necessarily aware of or are maybe actively trying to skirt some of the existing regulations and they think simply because it's a very novel technology that certain regulations don't apply** or they're outdated and therefore, won't affect them. Equally, some clinician researchers in the field are maybe **proceeding without necessarily contemplating how privacy law might interact with these technologies** or laws related to telecommunication technologies and so on." (ELPP_15) |
| | Need for Implementation Guidelines | "You **need governance pathways implemented with trust**, where the boundaries are clearly defined and operationalized. You can't simply ask the clinicians to use these tools out of the blue without these guidelines.**" (C_09) |
| **Unclear / Insufficient Regulatory Frameworks** | Regulatory Grey Zone | "So, one challenge that's coming up in the mental health field is where apps are pitched as being about, well-being rather than necessarily treating a mental health condition. An app might be doing very similar things, tracking the same data, continuous monitoring, and potentially suggesting certain things that might be going on. But **because it's built in terms of wellness, then that takes it outside of the governance regime of medical devices.**" (ELPP_15) |
| | | "Yeah, I mean, I think one important question to ask is, **is this thing offering me medical advice? Or is this thing practicing medicine?** And if the answer is yes, then it should have the requisite regulatory approvals to do so. And **if not, then it should be very, very clear that the clinician is in the loop...**I think **that's where any company maybe could get in trouble in that gray area.** There is a pretty bright line between practicing medicine and not. **You want to be very clear with your users, and as a user you want to be very clear what's happening here.**" (D_09) |
| | Oversight for Passive & Contextual Data Collection | "I think the second, so one is **regulation of contextual data in addition to medical data.** ... The problem is those are often not being mediated as medical applications. Those are **commercial applications that are collecting that data with little regulatory oversight.** So I'm very concerned about that." (D_19) |
| **Responsibility for Ethical Tech Development and Compliance** | Balancing Innovation with Regulation | "**Feasibility, it's super demanding. Something extra. [**Many developers are] saying 'We're already taking so many things into account and we're pretty willing to take ethics into account,'... But then **you come up with this [ethical] framework** which is... super specific and really demanding... And so they [are] **not so keen on taking on that extra work...**" (ELPP_08) |
| | Deploying Proprietary vs. Open-Source Algorithms | "I would say that **my concern,** looking around my [the tech development] industry, **is the blase attitudes that folks have when deploying proprietary algorithms for clinical use cases.** Because **we don't want our care system or clinical research practices to be in a place where they are dependent on a private algorithm to give access to care.** We've made that mistake in the past... being dependent on organizations like that... and so I wish that there were more checks and balances, or an awareness on that." (D_13) |
| | Unclear Liability | "**Are [patients] responsible** for seeing this alert and communicating? Is it internet-connected and alerting the clinical team? If [so], is it only for research purposes, and therefore you [the patient] actually have to pick up the phone and call your doctor? **Or does it create this false safety net that you think that your doctor is monitoring you because you have this wearable on you, but actually there's nothing that is alerting the doctor** to act on your behalf? So, I think **making sure that it's an inclusive approach and communicating to a patient how it works and what they're responsible for, and what the clinical team is responsible for, would be incredibly important."** (ELPP_19) |
| | | "**What's different here is the passive track...** [that] tracking happening in someone's personal and private life **outside of a clinical encounter...** I think there are **lots of open questions there about what the clinical relationship looks like** when that's happening, and **where the duty of care extends to.**" (ELPP_16) |
| **Need for Stakeholder Involvement** | Accounting for Lived Experience | "Any project that's **trying to develop a technology for clinical care should absolutely have... at least one person, if not multiple people, with lived experience.** If [it's] for depression and anxiety, well, you need people who've experienced depression and anxiety **involved as co-leads on the project.** ...A lot of tech companies and organizations are developing technologies because there's money in it, not necessarily because they actually want to solve a problem or care about the end, **the person who it's going to impact. [Those] people have alternative understandings of their symptomology, which aren't medical understandings. And they're valuable... Those kinds of understandings need to be incorporated into the technologies."** (ELPP_16) |
| | Need for Interdisciplinary Collaboration | "I'm working on teams with machine learning experts, with clinicians, with clinician researchers, with stakeholder groups, et cetera... general population consumers. One of the issues... is you get people who are really well-versed in this area. You get people who know the technology and are quite protective of their understanding of the technology. I see it happen all the time: You **get in meetings and the person who talks really well about the technology basically runs the show.**" (ELPP_14) |
| | Understanding How & Whether CP Matters | "**What is the clinical meaningfulness, even if we know that, 'Yeah, this is accurate.'** Those are some of the questions that come to my mind." (C_20) |

| | | "I think that is a con as well... that **a lot of this tech is developed without really involving users as to what kind of features are meaningful to them.** As far as I know, they're very rarely involved in this kind of research. **What actually matters to people?** What do they want monitored? Do they want these technologies? **Does constant monitoring in a sense,** make you feel better? What if you have health anxiety? ...Maybe you actually don't need... Maybe that is actually detrimental to you.." (ELPP_01) |

### *Data Privacy and Protection*

**Consent & Awareness.** Patients described anxiety over unwanted or unintended disclosure of intimate behavioral and physiological data, noting that continuous collection can feel like a privacy breach. Other stakeholder types likewise questioned the appropriateness of capturing real-time location or mental health indicators, characterizing it as invasive and, in some cases, "creepy." This unease was compounded by an awareness that elements of coercion may enter in play: individuals could feel pressured to share their data so as not to jeopardize access to healthcare services. Compounding these worries, stakeholders reported that explanations of data practices are often obscure, leaving patients unaware or uncertain about exactly what they are consenting to, who might have data access, what inferences could emerge from their data, and what kinds of feedback to expect. They may thus be ill equipped to make informed decisions around engaging with these CP tools, or what types of feedback to receive or decline (e.g., "right not to know").

Many participants, especially researchers, clinicians and ethics scholars, criticized current informed-consent practices as outdated and one-dimensional. Many acknowledged that patients typically encounter a single form at the outset of care (broad consent), without fully understanding the breadth of data collected or the myriad ways it might later be used. Certain respondents urged a shift toward dynamic consent models, in which patients receive clear, ongoing explanations and can granularly and dynamically opt in or out of specific data uses. They emphasized that such processes that view consent as an evolving conversation are better suited to the continuous, ecological monitoring characteristic of CP approaches.

**Secondary Use & Misuses.** Many patients and caregivers reported being comfortable with primary clinical uses of CP data but expressed concern about secondary applications and potential misuses. Stakeholders across groups noted that, without clear legal protections, patient information could be repurposed for discriminatory profiling or accessed by commercial actors, with existing regulations offering little specification how to manage these downstream uses. They argued that the commodification and monetization of personal behavioral and physiological data, in the absence of robust data protection frameworks, could deteriorate patient and caregiver trust in clinicians and healthcare systems and discourage future participation in digital health programs.

**Monitoring & Surveillance.** Stakeholders also observed that when individuals feel monitored rather than supported, they may withhold information, worry about data misuse, and question their providers' trustworthiness. This concern may be particularly relevant for vulnerable populations, such as people experiencing psychosis, who may perceive passive monitoring as surveillance, and older adults who may have difficulty using wearables and apps, highlighting the need for adaptive protocols, additional safeguards, and alternative engagement strategies that respect each patient's autonomy and comfort. Stakeholders noted that passive monitoring can shift the experience from feeling supported to feeling observed, leading individuals to withhold information, worry about how their data might be used, and question providers' trustworthiness. This effect may be especially pronounced among vulnerable groups; for example, people experiencing psychosis may interpret continuous tracking as intrusive surveillance, or members of historically exploited populations may hold significant reservations.

| Table 7. Data Privacy & Protection | | |
|---|---|---|
| Themes | | Illustrative Quotes |
| **Consent & Awareness** | Unwanted / Unintended Disclosure | "The biggest concern is just privacy... **it is going to collect stuff that you may not want another person to know, and if you don't realize all of that or you're not thinking all that through, it could be upsetting**...I mean, most of us have, our phones are [already] tracking us with all kinds of health data app." (C_12) |
| | | "[It's] **a big thing on your privacy**, then you're sharing a lot so **there might be times where, 'Oh, I don't really want to share things right now.'**" (P_16) |
| | | "**People have a right to privacy, and there's also just a limiting of agency in what you choose to share with your doctor or your provider.** Even **in clinical relationships or therapeutic relationships, the patient always has agency to talk or not talk about whatever they want, and share or disclose whatever they want. And that is also a meaningful process to allow patients to have agency in sharing that.**" (C_13) |
| | | "Well that **seems clinically very awkward. And uncomfortable. I think if a patient doesn't want to share something, there's probably a reason. And does that then put both the patient and the clinician in a weird position?** It's almost like when a parent tells you something about the kid but then says, don't tell the kid that I told you this. … **If it was sort of non-voluntarily shared, that seems kind of not ideal.**" (C_15) |
| | | "**People may say, 'Okay, I never consented my data to be screened for that…'**" (C_09) |
| | | "It's a little **bit creepy... Would I want all that information?** Generally, [we] have an expectation in life that certain things we can keep to ourselves and certain things are apparent to others. And **the whole idea behind this technology is, 'Hey, you can tell us things without actually telling us things. So we'll know things about you just by collecting this multitude of information about your facial expressions or your heart rate or things like that.' So there's that little bit of it that feels a little bit creepy.**" (C_14) |
| | Appropriateness | "I think there's a question about **whether we should be doing that, whether that is an appropriate way to integrate these kinds of tools into clinical care** when they're going to be tracking people in their homes, in their private lives, **revealing information that is sensitive and completely not health-related to their clinician...**" (ELPP_16) |
| | | There is **quite a distinct ickiness factor to being able to look that deeply into someone's personal life.** As a clinician, when I've had the opportunity to work with data like that or look at data like that, **even when it's de-identified, it's quite uncomfortable.**" (C_04) |
| | Coercion | "We almost **moved into this very coercive mode where** the alternatives, if you don't have these digital tools in some of the austere settings we're doing work right now, **the alternative is basically nothing.** So, what you're saying is like, **'I will give you this care or do this research and we will give you all of these things, but you have to give up all of your data.'** There is no alternative here." (C_07) |
| | | "It's like, **'Give us all this information and we'll provide you really good healthcare, but you can opt out. But just understand that we might not give you the healthcare that you need.'** There's not actually options there." (ELPP_14) |
| | Patient Communication & Understanding | "The other thing is about consent... **how would we set up consent for this information to be communicated to us?** Especially with this passive data collection..." (C_4) |
| | | "And of course, it **might differ for what kind of disease or condition you're experiencing**, but I think in lots of cases **people are quite vulnerable and it's really questionable whether they truly understand what the technology is doing, how it functions, what kind of tools it's using, what the end purposes is.** And that could then take on specific forms that are actually not beneficial for the patient." (ELPP_08) |
| | Perceived vs. Actual Data Sensitivity | "Most **people are not concerned about something like accelerometer data or things of that sort. [But] All those studies have demonstrated that you *can* identify the person... The perceived risk of accelerometer data is extremely low. But, that's only *perceived* risk. The actual risk is actually higher for that category.** For GPS, people are much more aware of that risk. But, again, **if you have a Facebook account, you already incur that risk of privacy loss,** specifically related to where you are... The resolution might be plus minus 100 meters, 200 meters, as opposed to GPS which is plus minus three or four meters. But, still, there's location there." (D_06) |
| | Patient Right "Not to Know" | "But if they **learn something about themselves they didn't want to know, that's hard.**" (ELPP_03); "Yeah, exactly. That's definitely an ethical dilemma. We have gene tests that do the same thing now. You can test someone to see if they have a gene for Alzheimer's disease or Huntington's disease before they ever developed symptoms of that and before you would even know. I think it's a similar thing, but there **are ethical implications to knowing that information and whether or not that information gets shared with patients and whether they want to know that information.**" (C_14) |
| **Secondary Use and Misuses** | Uncertain Secondary Uses | "I have no problem whatsoever using an app where I'm supposed to be clicking: 'I feel like I'm an eight out of 10 in terms of anxiety today' or letting it track my heart rate, letting it track my gait... **The GPS stuff seems like it starts to cross some lines too and I would just have a lot of concerns about how that data could be used....**" (CG_12) |
| | Discrimination Potential | **"Who knows how corporations may use that [CP data] in different ways to flag or discriminate against your child later on based on this kind of information, so that would be a concern, especially in the workplace."** (CG_02) |
| | | "Companies are sharing and using data for purposes other than why the data was initially collected... **If you really don't want your data used for other purposes, then there's not a lot of protections that exist right now.**" (C_14) |
| | Lack of Existing Data Protections | "**The more I learn about how the regulation works, what law governs what data can move and who can share what with [who]**... Every entity can be fully lawful, law-abiding and then still... **the framework, regulatory framework leaves holes, leaks for people to exploit.**" (D_18) |
| | Data Ownership& Control | "**This information [should be] kept exclusively for the treatment purposes of the pathology the patient comes in to treat..patient data should be the property of the patient,** and **only used within the realm that they have agreed upon.**" (C_10) |
| | | "We are selling out our healthcare data to groups like Google, Amazon, etc. **We are setting up our healthcare systems in a situation that healthcare professionals can no longer control their own data…**" (ELPP_14) |
| | Commercial Influence | "I think that as healthcare systems begin to act like corporations, the **corporatization of healthcare data will mean that monetizing facial expression data, location data, mood data, wearables, is going to be an incredible risk for healthcare systems.** So I'm very worried that healthcare systems have to be reinforced socially as having a **fiduciary duty to patients as patients, rather than... to stakeholders.**" (D_19) |
| | | "If my doctor had told me a year ago, I really want you to sleep better, so try this out," and I did it for a while and I thought it was only my doctor looking at it. [That becomes] very different, if I **could imagine ads being shown to me based on maybe Instagram somehow has access to my Oura data** on my phone and **can see that I'm acting manic, and they try and sell me specific things because of how my heart has behaved in the last week.** The crossing of the streams... thinking about what the potential benefits and harms are, that'd be the thing that would concern me... **The corporate environments, where the machinery that turns data into money, the people in those rooms, they don't live by [the] same rules [as us].**" (D_08) |
| **Monitoring & Surveillance** | Can Exacerbate Patient Distrust | "I think that some of the risks that I'm concerned about are how it would negatively impact the public's stigma towards mental health care. I think that it could dissuade people from engaging in that. **It feels like policing, which is not great, and I think for marginalized communities, extra not great...** I haven't really seen how this is applicable to kids, but **people already perceive me as policing them because I'm a mandated reporter,** so there's that." (C_13) |
| | | "And a reasonable response might be, 'Well, this is **just another form of surveillance. Why do they do this?**' And especially, again, in **people who might be vulnerable with mental health issues, the last thing you want to do is give someone something that they think is increasing surveillance if they're having some issues around surveillance.**" (ELPP_14) |

*Patient Impacts and Harms*

Stakeholders highlighted numerous ways that the above concerns may translate into direct or indirect harms for patients:

**Harms Due to Inaccurate or Premature Diagnoses.** Stakeholders from all groups cautioned that algorithmic assessments delivered without sufficient clinical context can trigger a cascade of inappropriate interventions. They warned that acting on false positives or early "flags" could expose patients to unnecessary tests, treatments, or stigma long before a human expert has had a chance to validate the finding. They also pointed out the negative impacts on patients when algorithmic conclusions may differ from patients' own perceptions and experiences, creating conflict with no clear resolution protocols.

**Diminished Human Connection in Healthcare**. A recurring theme, particularly among clinicians and patients, was the breakdown of human connection in healthcare. Many stakeholders noted that an over-reliance on data-driven CP tools could transform healthcare into a more transactional and less empathetic process. Clinicians especially highlighted the importance of maintaining therapeutic relationships characterized by respect, empathy and alliance, warning that digital tools, while potentially efficient, could diminish the "human touch" that is central to healing. Many patients and caregivers echoed this concern, fearing that healthcare interactions could become increasingly impersonal. Scholars and clinicians discussed the potential for digital health tools to contribute to epistemic injustice, whereby patients' lived experiences might be undervalued in favor of data-driven assessments. Some stakeholders worried that the emphasis on objective data could lead to the dismissal of patients' subjective experiences, particularly in complex areas like mental health where self-reports from patients already receive high levels of scrutiny. Such dismissals, they argued, could result in a loss of patient autonomy and a dehumanization of care, particularly if clinicians and patients allow algorithmic inferences to play an increasingly larger role relative to human judgment in decision-making.

**Responsibility Shifts and "Empowerment" Pitfalls.** Another significant concern raised by clinicians was the shifting of responsibility from healthcare providers to patients. With the increasing use of digital tools to monitor and manage health, patients are often expected to take on a greater role in their own care. While some viewed this as empowering, many clinicians worried that it could overwhelm patients, particularly those who might not have the skills, knowledge or interest to interpret data feedback. This shift could lead to feelings of confusion and stress among patients.

Ethics scholars also highlighted that while the rhetoric of "empowerment" is often used to promote these tools, it effectively pushes responsibility onto individuals, especially those with greater resources, while leaving vulnerable populations with insufficient mechanisms to address complex health inequalities. They pointed out that this shift not only burdens patients with the expectation of managing their health independently but also leads to blame when improvements do not occur, potentially exacerbating feelings of shame or anxiety. Certain ethics and policy scholars argued that this trend is reinforced by the technology sector's view of patients as consumers, rather than individuals needing care, framing health management as an individual, rather than collective, responsibility.

Additionally, clinicians noted the risk of patients deferring responsibility to technology, such as smartphones, under the assumption that these tools will manage their health, which can diminish active patient involvement in care. Patients who come to understand that their devices may "speak" for them may be less inclined (and eventually less able) to reflect on and articulate their experiences and behavioral patterns on their own.

**Access Inequities and Disproportionate Burdens to Vulnerable Populations.** Clinicians and scholars, primarily, voiced further concerns about the potential of CP tools to exacerbate inequities and disproportionately burden vulnerable populations. Scholars emphasized that marginalized groups, including those experiencing poverty, homelessness, and other forms of marginalization, may be excluded from the benefits of these technologies due to lack of access or capacity. For example, individuals without extensive access to or familiarity with technology might struggle to effectively use or trust these technologies, potentially limiting the potential benefits they may confer, and biasing training data sets in ways that perpetuate harmful biases and further exacerbate inequities.

Further, caregivers and ethicists, in particular, raised significant reservations around CP tools being leveraged or co-opted for surveillance, especially of communities with a history of being monitored, such as psychiatric and vulnerable groups. Pressured consent emerged as another concern, particularly for individuals in lower social positions who might feel compelled to use these tools despite discomfort or uncertainty. Finally, the risk of involuntary monitoring or detention was highlighted, with concerns that misdiagnoses or inaccurate data could lead to wrongful decisions, severely impacting individuals' rights and treatment.

**Threats to Privacy and Self-Determination.** Stakeholders from all groups voiced alarms about threats to privacy and autonomy posed by digital health tools. They expressed concern about the potential misuse of sensitive health data and the lack of transparency in how this information is collected and used. Scholars emphasized the need for stronger regulatory frameworks to ensure that patients' privacy is protected and that they have control over their personal health data. The feared that without adequate safeguards, the widespread adoption of these technologies could lead to breaches of trust and unauthorized access to sensitive information.

Clinicians pointed out that some patient populations are likely to be more disproportionately impacted by these concerns than others and may require especially robust clinical justifications and potentially enhanced protections or alternative approaches for using CP in ways that will benefit their care and ensure their rights to self-determination and discrimination avoidance.

**Epistemic Injustice and De-prioritization of Patient Voices.** Stakeholders cautioned that CP tools risk sidelining patients' own experiences by privileging algorithmic inferences over first-person testimony. Ethics scholars noted that even highly accurate systems can generate outputs that contradict a patient's self-knowledge, potentially leading clinicians to discount lived perceptions- and destabilize trust. Caregivers emphasized that real-time observations – such as a parent's instinct about a child's well-being – must carry equal or greater weight than sensor data to avoid silencing those closest to the patient.

**Overemphasis on Self--Optimization.** Experts warned that voluntary self-tracking can evolve into a cultural expectation, mirroring how smartphones became indispensable. What begins as clinically guided monitoring risks morphing into relentless personal optimization, pressuring individuals to engage in continuous self-surveillance. Stakeholders argued that blurring the line between medical indication- and -consumer-driven- tracking reduces complex human experiences to data points and undermines broader notions of wellbeing that cannot be quantified.

| Table 8. Patient Impacts and Harms | |
|---|---|
| Themes | Illustrative Quotes |
| **Harms Due to Inaccurate or Premature Diagnoses** | "I guess what's scary about it is I don't know **what happens when the algorithm gets it wrong** and somebody is *not* going into a depressive episode and they're called in and they're like, I'm fine. I don't need any intervention. Or **how will they feel about that**?" (D_10) |

| | | |
|---|---|---|
| | | **"What if, based solely on this computer technology, some kind of medication is prescribed and maybe the computer got it wrong,** or just something wasn't right, and then they were taking the wrong kind of medicine? That, I could see being an issue." (CG_20) |
| | | "At this stage, I'm not really sure there are many (CP models) that are clinically validated outside of specific controlled environments. And even if they were to be reliable in identifying things that do seem to correlate with what we know as depression or schizophrenia or whatever, **there are going to be instances where they get it wrong and then decisions are going to be made that have very, very serious implications for people."** (ELPP_16) |
| **Diminished Human Connection in Healthcare** | | "What we tend to forget or maybe overlook is quite often in mental health, the **human contact is really, really important.** So **if we're doing anything that, in fact, isolates people from their support networks or makes them think that they should be perhaps using this technology instead of seeking assistance with their human networks**… that can be **really damaging and really dangerous for people with mental health issues... the care relationship is really important in healthcare,** incredibly important in mental health and psychiatric health. That relationship, there's that **interdependent relationship. We rely on each other."** (ELPP_14) |
| | | ...I think **there's relationships that are formed with your doctor, and I just don't see that computers can ever replace that.** I hope they don't... but **if we get to a place where they do, I think that mental health in kids and adults will be worse off altogether because of the lack of value on relationships,** not just with your healthcare provider, but in general. (CG_17) |
| | | I think **there's a lot lost in terms of the physician or therapist-patient relationship...** So it used to be we would sit face to face and talk. Now... there's a computer there that's constantly being typed on and there's not a lot of eye contact... **that does concern me, that we're going to be more and more drawn into a virtual meta world rather than being in the real world with each other.** So I would want to see these used in conjunction with real human interactions... (C_03) |
| **Responsibility Shifts and "Empowerment" Pitfalls** | Patient Disempowerment | "We hear a lot of rhetoric, which is actually, I think, quite damaging. … this notion of empowerment [with digital health technologies]. … Really, what's actually happening is **we're pushing responsibility away from the state onto individuals.** People who are advantaged might be able to take charge and respond to that. A lot of people who need care aren't in the position to do that. So what we end up doing is we push blame. **Rather than making people empowered, we actually make them responsible, and then we blame them for their health inequalities and their health issues...** And then you get individuals who don't see improvements. It can **actually then worsen their self-image** where it's like, 'Oh, gee, I've got all this support and I still can't sort it out. Everyone else is. **I've got this tech. Why can't I get this? I'm empowered.** This is now my responsibility. **Why can't I do something about this?'"** (ELPP_14) |
| | | "I mean, **I feel apprehensive about the term empowerment** because immediately I ask, well, **who's saying someone is going to be empowered? Is it the person describing their own situation?** Or is it someone projecting that onto others? Empowerment is a word that seems to have been floating around mental health policy and probably other health policies since the 1990s... And was **part of a much broader trend towards transitioning from seeing a person as a patient to seeing them as a consumer of services**... And I think empowerment is something that has just continued through mental health services and other areas of health, and is definitely being taken on by the technology sector, because it is exactly **seen as an individualizing responsibility passed on to the person who's reframed as... a platform user. I think that narrative can hide a lot of those things...**" (ELPP_15) |
| | Deferral to Technology | "[The CP] system **might be doing it for them**... 'I **don't have to worry about that because my iPhone is doing it for me**' or whatever." (C_15) |
| **Access Inequities and Disproportionate Burdens** | Inequitable Access to Benefits | "There might be a large proportion of the population that benefits. … But we need to look at the boundary cases and the marginalized populations and the vulnerable and the extreme harm that we can do to them. … **Are we systemizing disadvantage? Is there systemic forms of discrimination? And with a lot of technology at the moment, the answer looks like it's going to be, yes, there is.** There's systemic disadvantage. Who's missing out? Okay, our aboriginal populations tend to be, our vulnerable groups, people who experience poverty, people who experience homelessness, our culturally diverse populations, our populations, LGBTQI+ plus populations. … So **these things really stack up** and become problematic." (ELPP_14) |
| | Surveillance and Policing | "We often forget that, certainly around psychiatric health, but **certainly around a lot of our marginalized vulnerable communities, they have a real fear of surveillance and interference. …** We've surveilled them and interfered in their lives so incredibly … So **if you give them a technology … a reasonable response might be, 'Well, this is just another form of surveillance.** Why do they do this?'" (ELPP_14); |
| | | **"Big Brother** is what I think of right away, and the **potential [data] misuse."** (CG_16) |
| | | "[Based on what] these technologies could determine or perceive... **would there then be automated [action]?** Just the implications of that. I think that that is something that I always have concerns about... **We're developing this technology for a hypothetical good in mind, but that technology could then be... leveraged into something that's more literal policing,** like would the police be dispatched if there's a mental health crisis? And knowing that **police involvement during mental health crises always goes badly for the folks who are experiencing mental health crises,** that's definitely a risk." (C_13) |
| | | "In terms of things to worry about, I think I probably worry about the idea of, really, persistent surveillance … **that's problematic for me that they turn into a tool of surveillance...**" (ELPP_09) |
| | Harmful Biases | "Who has access to this tool?... To the extent that that's **going to create a bias in the data sets that we get and the people that have access to this type of care, those are our potential concerns as well.**" (ELPP_09) |
| | Pressured Consent | "That's an inherent problem in all method of tracking or surveillance. I always worry about, **there's people who inherently are less empowered than others in virtue of their social status** … might feel less comfortable saying no to something like this, and it might actually exacerbate their mental health... **Do they feel this obligation to really do this quite invasive thing that might make them feel less at ease than they already are?**" (ELPP_12) |
| | Involuntary Monitoring or Detention | "So, the idea of introducing these tools that **might be used at scale to diagnose people and then to make treatment decisions for people who might be subject to an involuntary detention** order is scary, I think, where we are not actually sure that the tools are tracking anything. … **there are going to be instances where they get it wrong and then decisions are going to be made that have very, very serious implications for people, for their rights, for their treatment …**" (ELPP_16). |
| **Threats to Privacy and Self-Determination*** | Discrimination | "If [CP data could] I don't know how to describe it, work as a bias. He's applying to college right now. **Is there a potential that something could exclude him from a scholarship or an activity?**" (CG_19); "...thinking from a bad movie plot, could it be that someone's like, well, **we're going to eradicate all gender dysphoria, so anyone who's gender, this gets spiked on their data, we're pulling all them** and we're going to whatever, whatever. That sort of thing. I don't know. I'm sure for anything that there's good things about, there are ways that not good people can make it bad." (C_15) |
| | | "Identity theft… that there would be **a clone of my daughter somewhere**…" (CG_16) |
| | | "There is a well-known phenomenon [in] diabetes called diabetes distress … **continuously being exposed to your data … can actually lead to** … your glucose is better managed, but now **you are psychologically stressed out** about it so it has other consequences. So you **may be solving one problem and creating another one.**" (D_18); |
| | Psychological Burdens & Preoccupation | "Maybe [my daughter] **would hyper-focus on the results**, like how people say don't Google your results because your eye itches and now you have cancer behind your optic nerve… So that might be something because she does **tend to hyper-focus and obsess sometimes.** So that would be a concern." (CG_05); |
| | | **I don't want [my daughter] to have to think about [data collection] on a regular basis** and know that, 'Oh my gosh, everything I'm feeling, everything I'm doing is being recorded somewhere.' I wouldn't want it to affect her." (CG_20) |
| | | "Folks would argue, 'Yeah, but it's all just more information. You can do with it what you will.' **But that's not psychologically how people work.** If you tell me you have an increased risk of this, it's not just information. Now, I'm someone who has this. And how do I go about my day? So **I** |
| | | **really worry about returning 'insights,'** A, before they're super validated, but also how you return them." (ELPP_12) |
| | | "The manner in which you deploy and develop products around those algorithms. My primary care doctor is annoyed at how often her **patients will reach out to her and say, 'Oh my God, my Apple Watch told me this.'** And that is a**n example of the clinician being berated with data that is a distraction to them and their efforts to provide good care to their patients."** (D_13) |
| | Altered Self-Perception & Behavior | "[Information about my feelings or behaviors being automatically collected in the background] **might make me kind of self-conscious** about it and I don't know, it **might affect the way I act because if I'm trying to act a certain way to act for these devices, it might be changing me, which I don't know if that's a good thing."** (P_14); |
| | | "So **if we're recording your voice...is it going to affect how you interact with other people because it's being recorded a certain way? What are the sequelae for your social and personal life going to be?**… There's the people who can always say, 'Well, I don't want to do this anymore,' and then there's those who feel beholden to it and will continue to trudge along. So, yeah, sure, **that's something that I do worry about with that kind of application."** (ELPP_12) |
| | | I think maybe even the emotions could be very out of context, and if it were in context, that would be way too invasive. So that that's an issue. Monitoring location, I really don't like that. Again, anything that would have to do with food intake and even exercise. **Just things that if she's aware is being gathered, I don't know if it would be honest [accurate], because I think it would change the way that she would act.** CG_07 |
| | | "We humans lie. I don't know how that would be taken into account… **If you might sense that, 'Oh, all of this has been picked up. Do I have to start acting a certain way? Do I need to start saying certain things?** Is my response truthful? Is it not?' And so forth." (CG_16) |
| | Internalization of CP Inferences | "But then I worry, my bigger concern is then taking that and then relaying back to the patients or the users … **It's one thing to see raw data, but once you do the insight thing and you put the stamp of scientific approval on it, that means a lot to people**. And so they see that and they say, 'Oh my God, I'm depressed.' Or, 'Oh my God, I have a sleep disorder,' or whatever. … And then they **potentially self-incorporate that into their identities**." (ELPP_12) |
| | Gamification Impacts on Addictive Behavior | Yeah, it depends a little bit again on our understanding of vulnerability and whether or not we also have a sense of who might be more susceptible to these types of, I would say, habit-forming technology. … But **let's say they're at the moment engineered to habitualizing people to using this technology. Even this would be problematic in the consumer domain.** My understanding is that at Stanford University, you can actually take an engineering course where you can learn how to make apps more... addictive, if you will." (ELPP_18) |
| | Feedback Impacts on Autonomy | "Should people generally consent to getting these kinds of prompts and recommendations that guide their behavior in daily life with the promise that they feel better after a while, or **does this amount to a paternalistic nudging approach where self-autonomy is compromised?**" (ELPP_18) |
| **Epistemic Injustice and Deprioritization of Patient Voices** | | "Yes, we should respect the lived experience, the client's subjective perspective, but **what were to happen if there was this hypothetical oracle, this oracular system that was almost perfect, and it spat out something that was contra the patient simply because the patient wasn't fully aware of something? … going against the patient may create some other ripple negative repercussions.**" (ELPP_13) |
| | | "At a time in history where I suppose mental health and disability activists and people with that lived experience are trying to push back to some extent on expert knowledge as the basis for decision-making about policies and programming that affects people with mental health conditions... in some ways, **computational technologies like digital phenotyping are really promoting a kind of expert monitoring, that has the potential to make claims about what is happening with a person's internal state that may be quite different from what that person themselves is experiencing... that might have an impact on then how that person self perceives or how others perceive them.**" (ELPP_15) |
| | | "**I wouldn't want a provider to be using that data to exclude what I'm telling them** there in that moment... [about something] you're very concerned [about]... Because **I think that a parent's observations should be every bit as important as the data, if not more** depending on what exactly it is we're talking about. (CG_12) |
| | | "One of my concerns would be how it would affect the doctor relying on, like if you were advocating for your child or if you were advocating for yourself, **I would worry that they would place more emphasis on the technology, the information, than they would on [your concerns]...** And I realize some people aren't good at communicating, but then sometimes we have these mom instincts that are stronger, I feel like, than anything really. (CG_17) |
| **Overemphasis on Self-Optimization** | | I understand self-optimization, that's just some people's personalities... If one or two people want to do that, that's fine. **My concern is that it becomes more culturally acceptable to the point where we'll be starting to expect this of people the way no one's allowed to not have a smartphone today, you can't exist in society and not have a smartphone,** it's just absurd, **to get to a point where we're just expecting of people to have that level of self-surveillance.** And I think that there's some losses there with respect to how we exist in the world, this constant quantifying of the self... **Using it for a clear, clinical indication, versus using it for a project of self-optimization, I think that's a distinction that probably also maps onto the difference between the direct-to-consumer and the medical setting.** (ELPP_12) |

*Philosophical Critiques of CP*

**CP Is Insufficient to Capture Emotional States.** Certain scholars cautioned that CP technologies cannot truly capture the rich complexity of human emotion. They argued that feelings are not reducible to physiological impulses or static signals, but instead unfold in nuanced, shifting patterns that resist algorithmic measurement.

**CP Cannot Infer Emotion via Behavior.** Relatedly, some highlighted CP tools cannot infer emotion via behavior. While sensors can record facial movements, voice acoustics, heart-rate fluctuations and other behavioral or physiological phenomena, some said that these outward markers do not reflect internal experience and always demand human interpretation. One scholar likened the need for interpretation to how a radiologist must interpret and contextualize an image.

**CP Algorithms Embed Human Biases.** Other participants underscored that, because CP algorithms inevitably embed human biases, they cannot be purely objective indicators of pathology. They pointed out that every algorithm is built on manually labeled data and thus carries forward the cultural assumptions and blind spots of its creators. They argued that dependence on pre-coded categories mask underlying prejudices by erroneously situating CP outputs as "objective".

**CP Inferences are Not More Valuable than Subjective Patient Insights.** Some scholars also challenged the over-prioritization of data over dialogue and insisted that personal narratives, rooted in lived, phenomenological experience, offer primary and indispensable insights into illness that digital metrics cannot replace. They contended that patient testimony must "stand on equal footing" with any algorithmic outputs.

**CP Reflects Techno-Solutionism.** Scholars also warned that addressing illness primarily through a technological lens is part of a larger misconception that technology may solve everything. They highlighted the importance of instead attending to the social, political, and cultural dimensions of health. These stakeholders argued that an overemphasis on what can be measured or automated can lead to healthcare interventions that are shaped around the capabilities of machines rather than the holistic needs of people.

| Table 9. Philosophical Critiques of CP | |
|---|---|
| Themes | Illustrative Quotes |
| **CP is Insufficient to Capture Emotional States** | "I **don't think that you can detect emotion in that kind of automatic way**... People have spent so much time trying to do it. There are quite good scholars of emotions... who argue persistently and consistently, that that's not how emotions work, that's not how affect [works]. **It is not something that is measurable purely by a physiological impulse**...So **if you think that these technologies are going to... finally solve the problem of subjectivity within mental health research, it's not going to.**" (ELPP_01) |
| | "I worry that **we're reducing something that's potentially very complex** and that can be shifting over time." (ELPP_12) |
| | "The h**eterogeneous nature of humans and human behavior [and] mental health is a lot more diverse, disparate, and complex** than measuring, perhaps, diabetes…" (ELPP_13) |
| **CP Cannot Infer Emotion via Behavior** | "I think that in another concern is that **what a computer can measure is not an internal state; but that's what an emotion is,** and so conflating the two [emotion vs. behavior]. **Your face is moving in this way, so this computer's going to assume what that means is you're feeling this way,** I think it's a concern... **There's going to need to be some interpretation of that data.** Much like a pathologist reads a ultrasound or an MRI or something, **there's still a human interpreting that data**, I think they'll always need to be, because t**hose two things are not the same, an internal state, an outward expression of it.** "(C_19) |
| | "Well, **what worries me right off the bat is how you described what these technologies do, because they don't do any of those things. No technology recognizes, interprets or processes affect.** What these technologies do is they detect movements, of a sort... physical signals that are derived from movements. Maybe they're movements of the face, maybe they're movements of the heart, maybe they are the acoustics of a voice, but **they're not recognizing anything that is related to affect.**" (ELPP_03) |
| **CP Algorithms Embed Human Biases** | "**You can only train an algorithm on data that is pre-coded by a human being**. Someone has to do that manual human labor first. And digital phenotyping, no matter how much they sort of claim, **they can't get away from that problem**." (ELPP_01) |
| **CP Inferences not More Valuable than Subjective, Patient Insights** | "**Why do we assume that this kind of data that we might collect is going to be more objective or more reliable than engaging with patients about the phenomenological experiences of their illness?** I think that's a bigger question. It's like, okay, well yeah, maybe the **data is helpful, but it's not necessarily better than what people actually tell us about their experiences.** They **should have equal footing** when we're trying to understand illness and health." (ELPP_16) |
| **CP Reflects (Erroneous) Techno-Solutionism** | "I certainly see there's a lot of potential for these technologies, but my worry is that we have a conversation led by a certain perspective… this idea of solutionism or technological solutionism… So that causes me a lot of concerns, because what we then do is **we start to shape these social, political, cultural healthcare questions based around a tech perspective. And what that does is it overlooks a lot of issues that we see.**" (ELPP_14) |

## Discussion

### *Corroborating Existing Recommendations*

Our investigation illuminates the broad and varied concerns held by diverse stakeholders – developers, clinicians, patients, caregivers, and ethics and policy experts – around the translation of CP into clinical care. Understanding and responding to these concerns is critical to design implementation strategies that augment, rather than compromise, patient-centered and humanistic care. Many themes echo decades-old critiques of data-centrism in medicine: CP is simply the latest iteration of putting ever-richer "deep data" streams at the center of care, now amplified by powerful AI and machine-learning analytics. Accordingly, stakeholders reiterate familiar mandates from the trustworthy-AI canon, e.g., explainability, interpretability, bias mitigation, fairness, transparency. The opaque, "black-box" nature of many proprietary CP algorithms further exacerbates these issues, leaving patients and caregivers without clear evidence of how inferences about mood, cognition, or behavior are generated. Respondents in our study call, as many others have before, for robust, domain-specific validation standards, enhanced algorithmic transparency, liability frameworks for errors, contestability mechanisms, and guidance on reliably interpreting CP outputs across diverse clinical settings. But these are not new imperatives, nor are they disputed: there is broad consensus around the need for trustworthy algorithms and humanistic care.

Likewise, the call for implementation frameworks that safeguard clinician judgment, patient agency, and the therapeutic alliance is familiar. Stakeholders warned that uncritical, algorithm-driven monitoring risks displacing empathic dialogue by elevating decontextualized or biased metrics above patients' own narratives, shifting the therapeutic focus from shared understanding to automated inference. These fears are most acute for CP systems that directly infer diagnosis (classification) and prognosis (prediction) but may be less prominent when CP is used to surface raw patterns—e.g., sleep or activity metrics—for human-guided interpretation. For example, rather than leaving an algorithm to characterize sleep patterns as pathological, clinicians could leverage a patient's baseline sleep data (juxtaposed with population benchmarks) to ask, "What's keeping you up at night?" and collaboratively co-identify what constitutes normal sleep for that individual in the context of work, family, or lifestyle factors. D'Alfonso and colleagues [9] have referred to this distinction as "manual" versus "AI-driven" use of CP, which highlights the degree of human involvement in data interpretations. At the time of writing, most CP tools are not yet robust enough to fully rely on AI-driven inferences and thus require a significant degree of human interpretation to be clinically useful. However, as we argue elsewhere [52], this will not always be the case, and – following the ascendant curve of AI in other domains – CP algorithms are likely to advance to the point of offering valid, accurate, patient-specific and trustworthy inferences. The need to establish humanistic approaches well in advance is a consensus goal.

*Novel Insights: The Importance of Context and Subjectivity*

Our respondents pointed out two fundamental considerations for effectively and humanely translating CP tools into care that have yet to be elaborated elsewhere: the importance of context and subjectivity in determining the clinical significance of CP outputs. Stakeholders across all groups stressed that observable behaviors – steps, voice tone, facial micro-movements – are clinically actionable only when clinicians understand what those behaviors mean for the individual who produces them and how the surrounding situation shapes that meaning.

This warning echoes the "theory of constructed emotion" advanced by Barrett et al. [53] and like-minded scholars [54-57] who reject the classical view that emotions are biologically hard-wired states expressed through universal behavioral markers. Instead, the brain constructs each feeling from past experience, cultural learning, and moment-to-moment interpretation; the same smile can signal joy, embarrassment, or compliance, depending on context [58, 59]. When CP systems infer affect solely from facial features, vocal prosody, heart-rate variability, or other external cues, they risk flattening this complexity into generic labels, an error that disproportionately misreads people from different cultures, age groups, or clinical presentations [60].

To counter such reductionism, future CP strategies must map subjective meaning and environmental context alongside sensor data. Technically, this entails pairing passive streams with structured self-report or ecological annotations that capture the patient's own interpretation of events and the situational factors at play. Operationally, it calls for structured conversations –from earliest visits throughout follow-up – that identify which symptoms most constrain a person's quality of life and how those symptoms might be detected digitally. The "Digital Measures That Matter to Patients" framework proposed by Manta and colleagues [61] offers concrete guidance here, linking *meaningful aspects of health* to sensor-derived concepts of interest, outcomes, and endpoints in a patient-centered hierarchy.

In practice, applying this framework would mean, for example, that a patient who values uninterrupted sleep over daytime mood stability might prioritize actigraphy-based sleep metrics, whereas another concerned about social withdrawal could ask the system to flag sustained drops in communication patterns. By weaving patient narratives and situational detail into metric selection and interpretation, clinicians can transform CP from a one-size-fits-all detector into a context-aware, individually tailored decision support tool, staying true to the subjective richness that stakeholders warn must never be lost.

### *A Prototype for Humanistic Care with CP*

To address these challenges, we introduce the concept of personalized roadmaps [62] for integrating CP into clinical care – a structured, co-designed plan that embeds humanistic values into every stage of digital phenotyping. Rather than treating data feedback as a series of discreet disclosures, personalized roadmaps are developed collaboratively by patients, caregivers, and clinician researchers at the point of consent. Together, they specify:

- **Which metrics** (e.g., activity patterns, speech markers, sleep variability) will be monitored and shared,

- **When and how** these data will be returned – whether in real time, during clinic visits, through periodic summaries, or some strategic (non-arbitrary) mix of approaches,

- **Thresholds for action**, delineating what combinations of signals should trigger outreach, referral, or adjustment of treatment,

- **Conflict resolution procedures** for managing epistemic conflicts when CP outputs diverge from a patient's self-report or a clinician's judgment.

This iterative framework balances patient agency with clinical and ethical guardrails, inviting patients to contribute lived knowledge (for instance, recognizing that reduced text messaging often precedes mood dips), while researchers share their clinical expertise, and both parties anticipate and come to shared understandings about how their perspectives may be enriched by predictive insights from CP data trends and inferences. This approach is based on a belief, articulated by others [49], that technology and humane care are not mutually exclusive, and in fact, can be symbiotic. The personalized roadmap is intended to foster that symbiosis, offer a living decision-support tool that aligns computational power with at least three operationalized, person-centered goals of care, including:

**Empowerment & Shared Decision Making.** By inviting patients to co-select which CP signals matter most and how they wish to receive feedback, roadmaps transform passive monitoring into an active partnership. This expands on Schmidt and D'Alfonso's finding that clinicians and clients value systems where patients can "switch off" sensors, control data sharing, and iteratively refine monitoring parameters [47]. Patients could collaboratively choreograph the timing, dose and content of feedback to align with their treatment goals. Embedding these choices upstream prevents downstream surprise or distress when digital inferences arise.

**Trust & Therapeutic Alliance.** Clear, co-crafted expectations—what will be returned, when, and under what conditions—mitigate nocebo effects and overreliance on opaque risk scores. As Nghiem et al. observed [46], passive patient-generated health data are most useful when presented at clinically meaningful moments rather than flooding clinicians in real time; roadmaps can prescribe that timing, ensuring data review occurs within empathetic, dialogic encounters rather than interrupting them.

**Ethical Transparency & Anticipation of Conflict.** Documenting both inclusion and exclusion of specific CP metrics is inspired by the "open notes" movement, giving patients insight into the analytic process and preserving their right to know what factors shape their treatment pathways, as well as their right "not to know" certain inferences that may be counterproductive towards clinical progress. Roadmaps also embed anticipatory strategies for epistemic conflicts; for example, if a wearable flags elevated stress when a patient reports feeling calm, a roadmap can offer co-identified strategies to guide the clinician and patient through a respectful dialogue about potential device errors, contextual factors, or unrecognized symptoms, rather than defaulting to algorithmic authority (or supremacy of patient report).

### *Innovating Consent for CP Approaches*

As CP technologies shift from the realms of clinical research into care, these roadmaps will support clinical teams with a fiduciary responsibility to educate patients about the anticipated benefits and risks and to transparently convey where impacts remain uncertain. Improving upon existing consent procedures should begin with an identification of patient and caregiver knowledge needs for informed consent. In a recent publication, we present results of an empirical, qualitative analysis [63] of perspectives among adolescent patients and their caregivers participating in clinical lab research entailing extensive collection CP data. Our findings demonstrated that they have information needs across seven key domains: (1) clinical utility and value; (2) evidence, explainability, evaluation and contestation; (3) accuracy and trustworthiness; (4) data security, privacy, and misuse; (5) patient consent, control, and autonomy; (6) physician-patient relationship; and (7) patient safety, well-being, and dignity. A separate analysis [40] found that most patients and caregivers view CP data as highly sensitive and expressed a reluctance to share these data beyond their clinical teams. While many participants expressed trust in existing data protections to protect CP data, they often misunderstood or overestimated the extent of protections like HIPAA to safeguard CP data. Based on these findings, we proposed five key strategies, including: (1) educating patients on the limitations of existing data protections; (2) conducting targeted research, including forensic analyses, into secondary data exchanges and identify privacy breaches or reidentification risks; (3) enacting regulations that mandate greater transparency in health data transactions; (4) implementing computational mechanisms, such as distributed ledger technologies, to enhance data traceability and auditability; and (5) adopting dynamic consent models that allow patients to continuously manage and update their consent preferences.

Other scholars have likewise argued that static, one-time signatures are inadequate for the continuous, highly contextual data streams generated by CP tools. A systematic review of ethical considerations for passive data sensing [64] proposes interactive informed consent interfaces that let participants add social annotations, "talkback" questions, and multimodal visual aids – features shown to improve comprehension and engagement [65, 66]. Others have called for consent models that are context-sensitive [67, 68], giving patients the ability to recalibrate permissions as circumstances change, and enabling built-in data expiration options, allowing individuals to set automatic sunset dates [69]. These consent innovations should be integrated into the personalized roadmap architecture to ensure that consent is not a static but an evolving agreement.

### *Operationalizing Humanistic Use of CP*

Most would agree that maintaining a sense of humanity in care is critical – and in fact, we already have a reasonably clear vision of what humanistic practice looks like, even if current systems fall short. Humanistic care is compassionate, respectful, and empathetic. It is also collaborative, culturally sensitive, and empowering. The formative research presented here corroborates a substantial body of prior work [70-72] showing how diverse stakeholders conceptualize and idealize humanistic care. In other words, further studies to delineate what constitutes humanistic practice and to demonstrate its benefits for patients, clinicians, and communities are no longer the priority; that foundational work has already been carried out. Instead, what is now required is rigorous, context-specific evidence for which CP integration strategies most effectively embody these established humanistic care ideals, i.e., which organizational policies, device design features, relational practices, and value-based attitudes to incorporate and which to eschew. We still lack evidence-based guidelines for integrating CP; and the only way to develop them is to investigate a wide spectrum of implementation contexts and determine which combinations of features produce desired outcomes, for which patients, and under what circumstances. Our analysis highlights several feature domains that require systematic evaluation:

- **Data handling:** collection methods, governance structures, and privacy safeguards
- **Feedback logistics:** cadence, routing, and escalation pathways
- **Patient support:** education, engagement, and shared-decision tools
- **Analytics:** modelling choices, interpretive aids, and decision-support mechanisms
- **Interface design:** usability, accessibility, and visualization elements
- **Workflow integration:** infrastructure requirements and task allocation
- **Clinician readiness:** training, supervision, and capacity building

Each domain contains multiple variables whose effects will differ by setting. Treating these variables as elements in a "constellation" and iteratively testing how their configurations shape clinical and humanistic outcomes will allow us to pinpoint the scenarios in which specific approaches add value and those in which they do not. Such empirical investigation may reveal that CP approaches are not for everyone, or every clinical scenario.

### Concluding Reflections

Integrating CP technologies into everyday clinical workflows surfaces specific tensions that can undermine even the most deeply held humanistic ideals. Countless forces compete with our ability or desire to deliver humanistic care. In the case of CP, one of the most pervasive is the shared conviction—among clinicians, patients, and caregivers alike—that data speak more objectively than lived experience. As our stakeholders warned, centering illness interpretations on digital signals risks recasting patients' stories through the lens of machine-generated feedback. Anthropologists describe this phenomenon as an "idiom": a culturally patterned mode of expression, whether verbal, behavioral, or somatic, through which distress or wellbeing is communicated in ways that reflect shared meaning, based on local beliefs and values. Classic idioms of distress—"heavy heart [73, 74]," "*ataque de nervios* [75]," or notions of hot–cold imbalance [76]—operate less as discrete biomedical signs than as symbolic languages that link individual suffering to broader cultural meanings, social relationships, and moral concerns. If data become the dominant idiom through which we express or even conceptualize illness, the less able we may become to recognize, convey and intervene in the complex multitude of factors influencing illness.

These idiomatic shifts pose far graver threats than concerns about false alarms, opaque metrics, or data privacy – issues that, while critically important, are largely tractable and already commanding intense scholarly and technical attention. By contrast, the deeper danger lies in narrowing our collective capacity to perceive human realities by privileging quantifiable signals over the nuanced psychosocial factors that shape how illness is understood and experienced. In this light, dehumanized care reflects not merely a violation of respect or rights, but a siphoning-off of human insight, potentially leading to an atrophy of clinicians' curiosity and compassion and of patients' ability to articulate their own experiences.

Ironically, this outcome runs counter to CP's original promise: to deliver objective, reliable windows into complex disease states and, in doing so, draw us nearer to the ground truths of human suffering. Data alone cannot constitute those truths. The critical question – one that our study helps to inform – is how to weave these deep data into care in ways that enhance, rather than diminish, the humanistic foundations of care.

## Acknowledgements

We would sincerely like to thank all of the clinicians, developers, ELPP scholars, patients and caregivers who took the time to interview with us and share their valuable perspectives. Additionally, we would like to thank the research coordinators from our "sister" study who helped recruit patients, including Rebecca Greenberg, Jessica Foy, and Yuen Yu.

This research was funded by the National Center for Advancing Translational Sciences (R01TR004243). Views expressed here are solely those of the authors and do not necessarily reflect the official policies of the National Institutes of Health (NIH) or the U.S. Government.

## Conflict of Interest

ES reports receiving research funding to his institution from the Ream Foundation, International OCD Foundation, and NIH. He was formerly a consultant for Brainsway and Biohaven Pharmaceuticals in the past 12°months. He owns stock less than $5000 in NView. He receives book royalties from Elsevier, Wiley, Oxford, American Psychological Association, Guildford, Springer, Routledge, and Jessica Kingsley.

The remaining authors declare that the research was conducted in the absence of any commercial or financial relationships that could be construed as a potential conflict of interest.